\documentstyle[12pt,epsfig,blois]{article}
\begin{document}

\def\Onow{\Omega_0}
\def\Lnow{\Lambda_0}
\def\kms{\,{\rm {km\, s^{-1}}}}
\def\msun{M_\odot}
\def\vcir{V_{\rm c}}
\def\plottwo#1#2{\centering \leavevmode
\epsfxsize=.45\columnwidth \epsfbox{#1} \hfil
\epsfxsize=.45\columnwidth \epsfbox{#2}}

\heading{THE FORMATION AND EVOLUTION OF DISK GALAXIES}

\author {Shude Mao, H.J. Mo, Simon D. M. White}
{Max-Planck-Institut f\"ur Astrophysik,
85748 Garching, Germany.} {\  }

\begin{bloisabstract}

We review some of our recent progress in modelling the
formation of disk galaxies in the framework of 
hierarchical structure formation. Our model is not only
consistent with the local observations, but also provides a good
description of disk evolution out to redshift $z \sim 1$. 
We use this model to interpret recent observational results
on Lyman-break galaxies (LBGs) at $z \approx 3$. Assuming 
that these galaxies are associated with the most massive halos 
and adopting an empirical law for their star formation rates, 
we find that many properties of the LBG population, 
including their correlations, sizes and kinematics can be accommodated 
in the model. 
\end{bloisabstract}

\section{Overview of disk formation}

In standard hierarchical models dissipationless 
dark matter aggregates into larger and larger clumps as 
gravitational instability amplifies weak initial density perturbations.
Gas associated with such dark halos cools 
and condenses within them, eventually forming 
the galaxies we observe today. 
These two aspects of galaxy formation 
are currently understood at different levels. 
The mass function, density profiles and angular- momentum
distribution of dark matter halos
are well understood from numerical simulations and from
analytical treatment; we take these as direct input in our modelling.
On the other hand, the gas assembly and the related 
star formation and feedback processes are
not understood in detail. For this uncertain process, we use an updated
version of the scheme proposed by Fall \& Efstathiou \cite{fal80}.
Briefly, after the initial protogalactic collapse the gas and dark
matter are assumed to be uniformly mixed in a virialized object with
density profile modelled by the NFW profile \cite{nfw}.
We assume that the gas component gradually settles into 
an exponential disk and that the gas conserves angular momentum 
during this process until it finally reaches rotational support.

This model reproduces the
properties of local disk galaxies quite well, including the shapes of
rotation curves, the size vs. circular velocity relation and, 
most importantly, the Tully-Fisher relation and its scatter, 
see \cite{mo98a}.

\section{Disk evolution out to redshift $\sim 1$}

\begin{figure}
\centerline{ \psfig{file=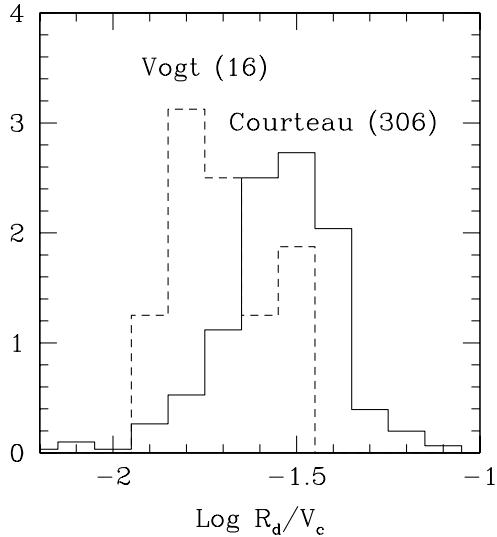,width=8cm} }
\caption{
Histograms of the ratio of disk scalelength to (maximum) circular
velocity of galaxies. The solid line is for the local sample from
\cite{cou96}, \cite{cou97} and has 306 galaxies.
The high-redshift sample is from \cite{vog97a},
\cite{vog97b} and includes 16 galaxies.
} 
\label{fig:fig1}
\end{figure}

The disk population is predicted to evolve. For example, the disk
scalelength is expected to change according to
\begin{equation}
R_d \propto \lambda V_c\frac{H_0}{H(z)}\,\,,
\label{eq:rd}
\end{equation}
where $R_d$ is the disk scale length, $V_c$ is the circular velocity of
the dark halo, and $H(z)$ and $H_0$ are the Hubble constants at 
redshift $z$ and $0$, respectively.
Since the Hubble constant increases with redshift, for given
circular velocity the disks are predicted to be smaller
at higher redshift. To single out the
redshift dependence, we use
the marginal distribution of $R_d/V_c$. A comparison of this quantity
between the local and high-redshift samples
reveals a size evolution of disk galaxies (see Fig. \ref{fig:fig1}). 
The amount is similar to the model prediction. 
More comparisons between theory
and observations can be found in \cite{mao98}.

As is clear from Fig. \ref{fig:fig1}, the high-$z$
disk sample is still quite small, and the
evolution out to $z \sim 1$ is quite modest. 
It is important to go to higher redshift in order
to probe the predicted evolution. 

\section{Application to Lyman-break galaxies}

The Lyman-break technique has been very successful
in discovering many galaxies at $z \sim 3$.
The properties of these galaxies are reviewed in detail by 
Steidel (these proceedings) and are therefore not repeated here. 
We assume that LBGs are the
central galaxies of the most massive dark halos present at $z\sim 3$. 
The gas in the halo is arrested either by its spin, or by fragmentation 
as it becomes self-gravitating. We use the empirical results of 
\cite{ken98}, calibrated by local galaxies,
to determine the star formation rates.
We identify LBGs as galaxies with the highest SFR and normalize
their number density to the observed value.

\begin{figure}
\epsfysize=8cm
\plottwo{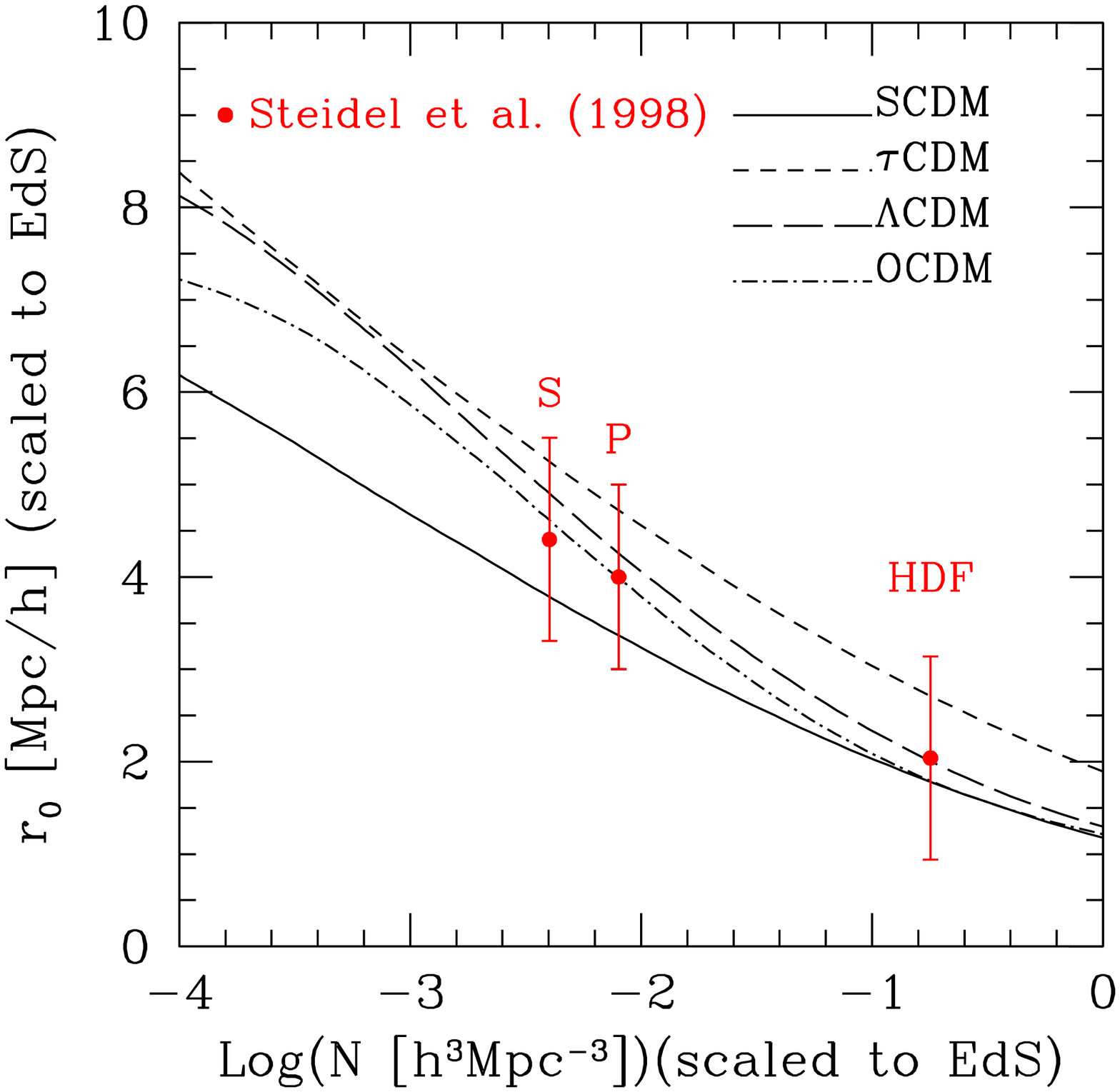}{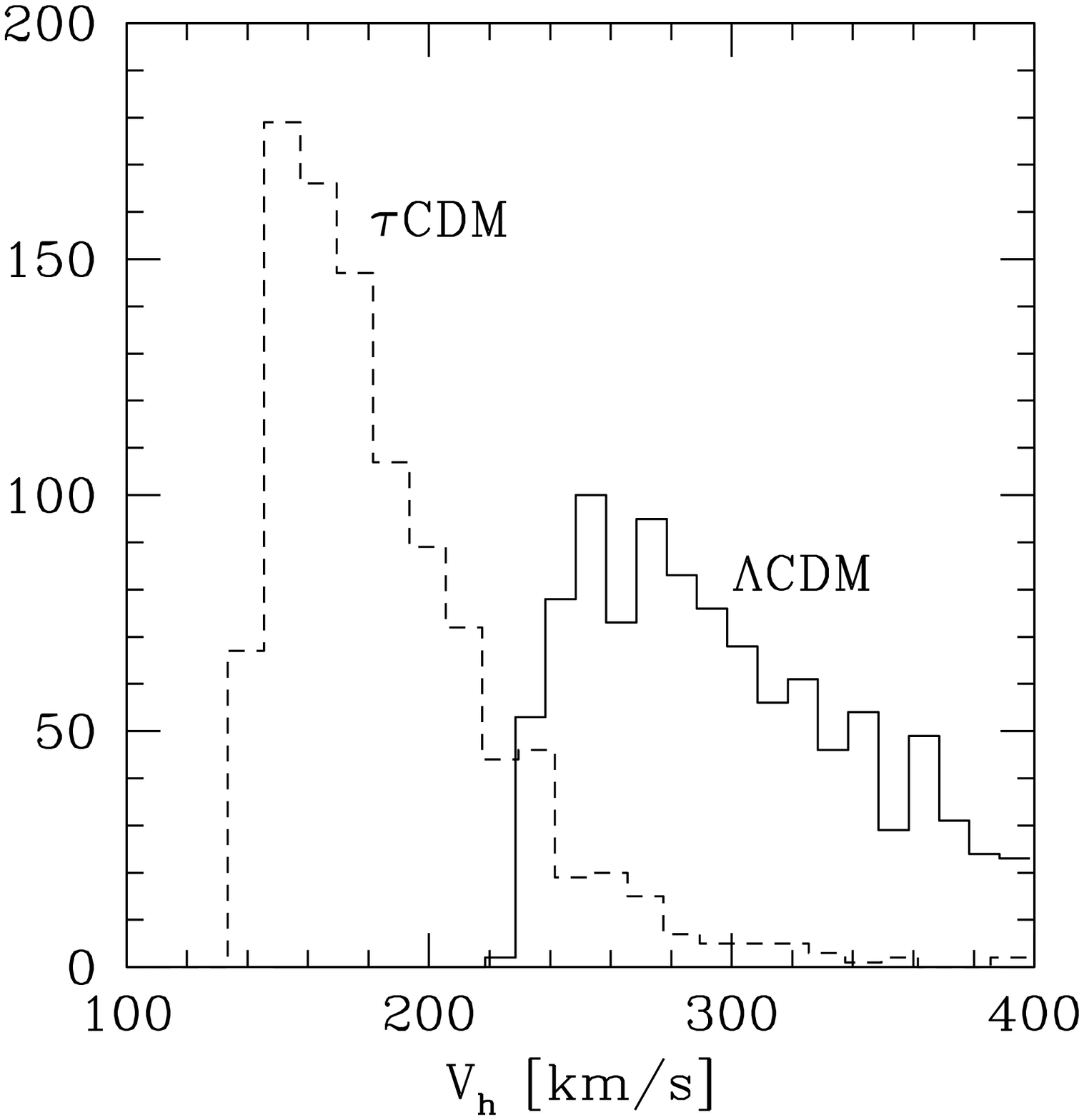}
\caption{The left panel shows correlation length
as a function of abundance. Both quantities have been scaled 
to the Einstein-de Sitter (EdS) cosmology by applying
appropriate correction factors.
Results are shown for four cosmogonies, as indicated
in the panel. Observational data for LBGs to three different
limiting magnitudes are taken from \cite{ste98}.
The right panel shows the circular velocity distribution for the
halos which host LBGs. 
}
\label{fig:fig2}
\end{figure}

The left panel of Fig.\,\ref{fig:fig2} shows the correlation length,
$r_0$, as a
function of abundances of LBGs, $N$, for four different cosmogonies. The
models are defined in \cite{mo98b}. Here, we only 
outline the two models for which some detailed results 
are summarized in the following.
1) $\tau$CDM:    $\Onow=1.0, \Lnow=0.0, h=0.5, \Gamma=0.2, \sigma_8=0.6$
and 2)
$\Lambda$CDM: $\Onow=0.3, \Lnow=0.7, h=0.7, \Gamma=0.2, \sigma_8=1.0$.
The model parameters have their usual meanings 
(see \cite{mo98b}). As we can see from Fig.\,\ref{fig:fig2},
all four models
reproduce the observed correlation strength well. The observed clustering
alone does not provide strong constraints on cosmological models.
To discriminate between cosmological models, information about the 
internal structures of LBGs is needed.

Fig.\,\ref{fig:fig2} (right panel)
shows the distribution of circular velocity 
for the LBG host halos. In $\Lambda$CDM, these
circular velocities are quite big, with
a median of about $290\kms$, corresponding to 
a total halo mass $M_{\rm h}\sim 1.0\times 10^{12}
h^{-1}\msun$. In this model, LBGs are indeed associated 
with massive dark halos. In contrast, the halo circular velocities in the 
$\tau$CDM model are much smaller. The median is now
about $180\kms$, corresponding to $M_{\rm h}\sim 
1.5\times 10^{11} h^{-1} \msun$. In this cosmogony, relatively few massive
halos can form before $z=3$, and one has to include lower mass systems
in order to match the observed number density of LBGs. 

\begin{figure}
\epsfysize=8cm
\plottwo{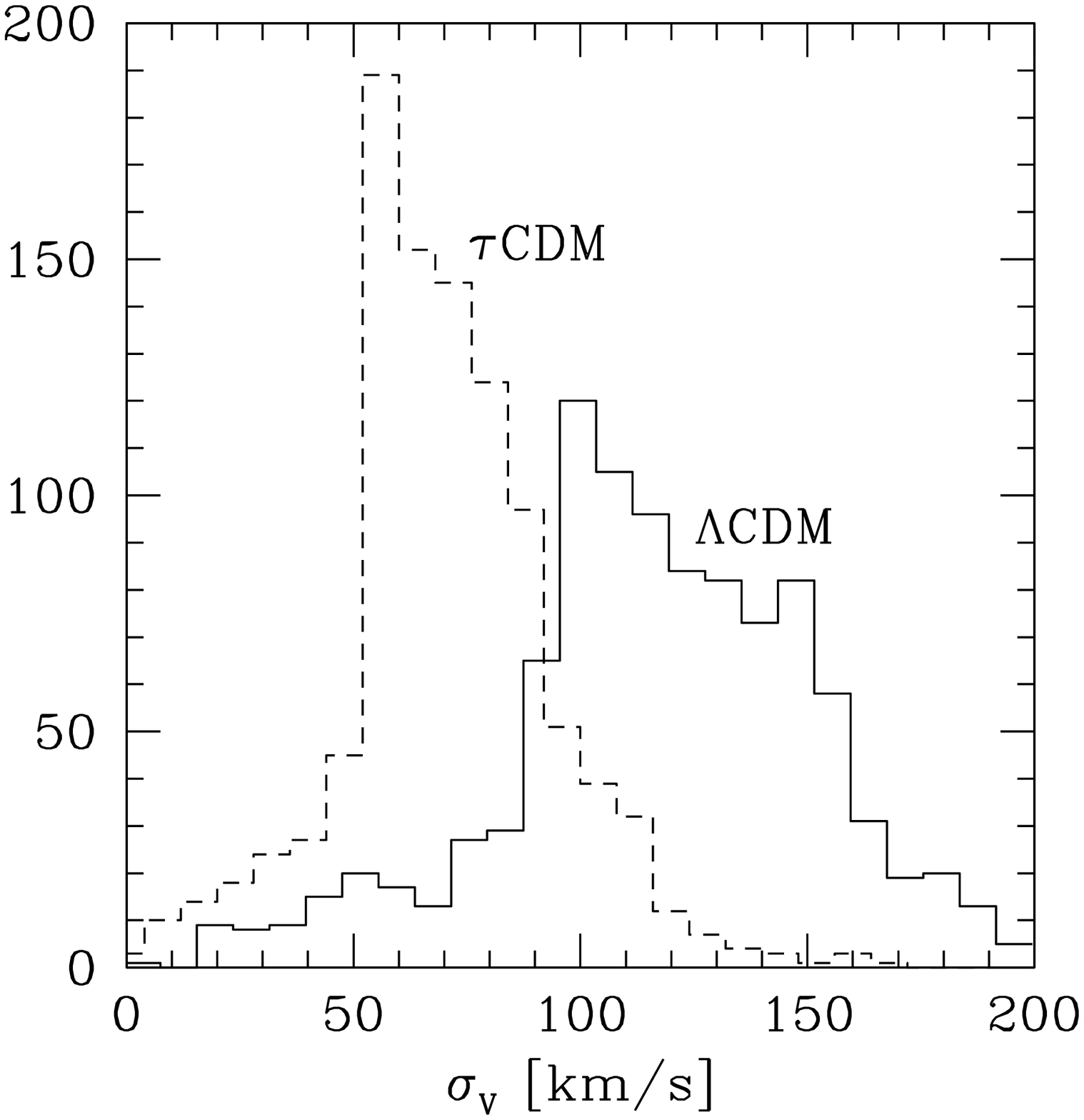}{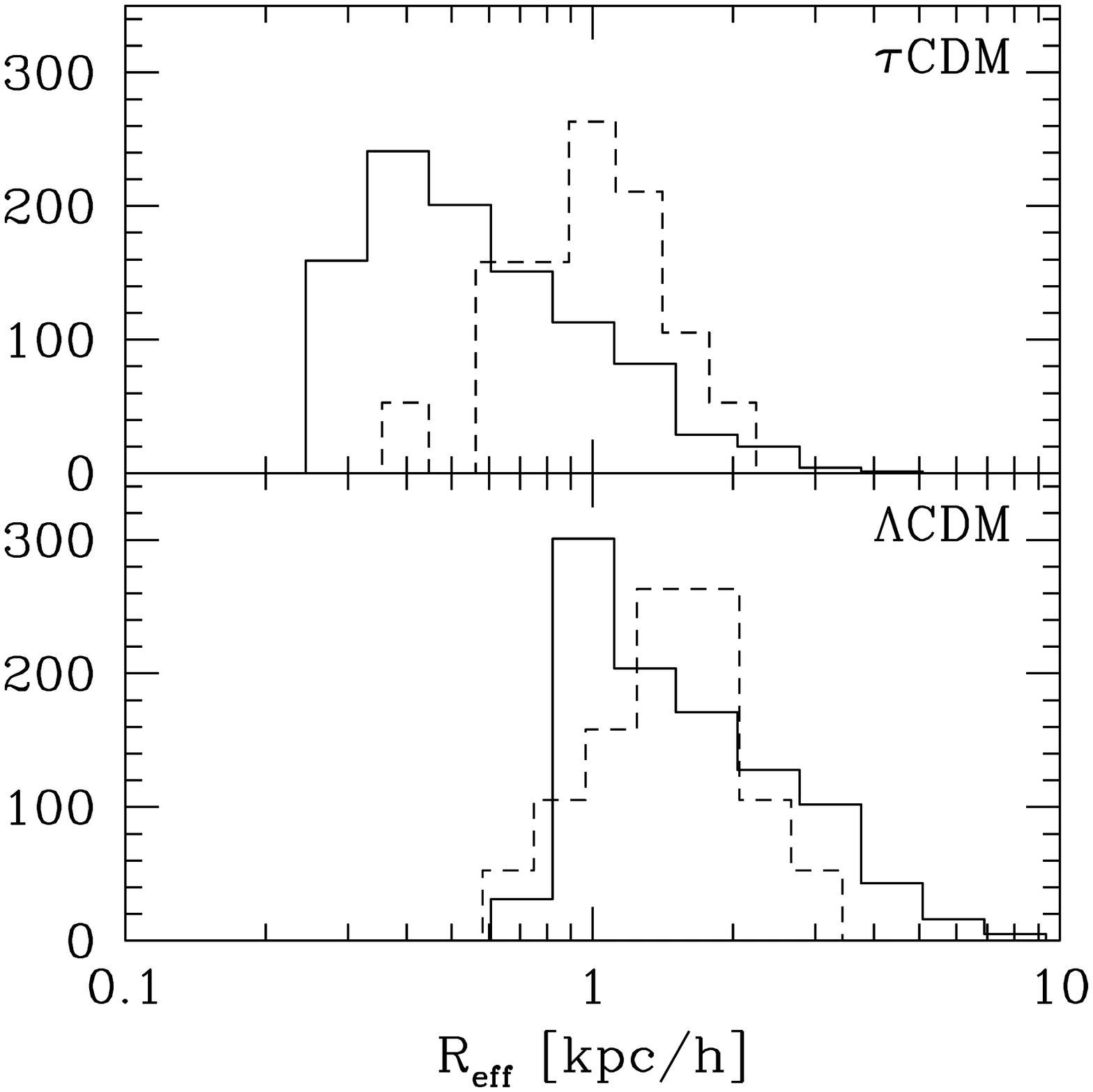}
\caption{
The left panel shows
the line-of-sight stellar velocity dispersion distribution of LBGs. 
Solid and dashed histograms are 
for $\Lambda$CDM and $\tau$CDM, respectively.
The right panel shows
the distribution of half-light radii, $R_{\rm eff}$, for LBGs. 
The solid histogram gives the model prediction and the
dashed histogram shows observational data.
}
\label{fig:fig3}
\end{figure}

Although our model predicts the circular velocities of 
LBG host halos to be quite big, the stellar velocity 
dispersions of the LBGs themselves are roughly a factor of 2 smaller
(cf. the left panel of Fig.\,\ref{fig:fig3}).
This is a result of a combination of projection effects with the fact
that the observed stellar distribution samples only the very central
regions of the halo potential well where disk rotation curve  
is still rising.

The right panel of Fig.\,\ref{fig:fig3}
shows the predicted distribution of half-light radius, $R_{\rm
eff}$, for the LBG population. The prediction closely matches the
observed sizes
for the $\Lambda$CDM model while the $\tau$CDM model seems to predict
sizes that are somewhat too small compared with the observations.

\section{Conclusions}

In conclusion, the formation and evolution of galactic disks 
can be well understood in the general framework of structure 
formation. The model we are studying seems to provide an
adequate description of local disk galaxies \cite{mo98a}
and of disk evolution out to $z \sim 1$ \cite{mao98}.
The structure and clustering properties of Lyman break
galaxies at $z \sim 3$ are also nicely explained
on the hypothesis that they are the high-redshift equivalents of local
bright disk galaxies \cite{mo98b}.

\vfill
\end{document}